\documentclass[english,superscriptaddress,preprint]{revtex4}
\usepackage[T1]{fontenc}
\usepackage[latin9]{inputenc}
\usepackage{textcomp}
\usepackage{graphicx}

\providecommand{\tabularnewline}{\\}

\usepackage{babel}

\begin{document}

\title{A Search for pair production of the LSP $\tilde{\nu_{\tau}}$ at
the CLIC via RPV Decays}

\author{M. Sahin}

\email{mehmet.sahin@usak.edu.tr}

\affiliation{Department of Physics, Usak University, Usak, Turkey.}

\author{A. B. Sener}

\email{absener@gazi.edu.tr}

\affiliation{Department of Physics, Gazi University, Ankara, Turkey.}

\author{S. Sultansoy}

\email{ssultansoy@etu.edu.tr}

\affiliation{Physics Division, TOBB University of Economics and Technology, Ankara,
Turkey and Institute of Physics, Academy of Sciences, Baku, Azerbaijan.}

\author{M. Yilmaz}

\email{metin@quark.fef.edu.tr}

\affiliation{Department of Physics, Gazi University, Ankara, Turkey.}
\begin{abstract}
In this work we consider pair production of LSP tau-sneutrinos at
the Compact Lineer Collider. We assume that tau-sneutrinos decays
in to e\textmu{} pair via RPV interactions. Backgroundless subprocess
$e{}^{-}e^{+}\rightarrow\tilde{\nu_{\tau}}\bar{\tilde{\nu_{\tau}}}\rightarrow\mu^{+}\mu^{+}e^{-}e^{-}(\mu^{-}\mu^{-}e^{+}e^{+})$
is analyzed in details. Achievable limits on $Br\,(\tilde{\nu}_{\tau}\rightarrow\mu e)$
at $3\sigma$ and $5\sigma$ CL are obtained depending on $\tilde{\nu_{\tau}}$
mass.
\end{abstract}
\maketitle
Supersymetry (SUSY) is one of the favorite candidates for the Beyond
the Standard Model (BSM) physics \cite{Haber}. For this reason, searching
for supersymetric particles forms an essential part of the LHC, as
well as future colliders, experimental programes. Searching strategy
for SUSY strongly depends on the lightest supersymetric particle (LSP).
As it was shown in \cite{Sultansoy}, the LEP data allows {}``right''
sneutrino to be the LSP. In addition, R-parity violation allowing
decays of LSP sneutrino to ordinary SM particles leads to potentially
rich phenomenology at high energy colliders \cite{Barbier,Dimopoulas,Barger1,Barger2,Sierra,Cakir1,Cakir2}.
R-parity is represented by $R=(-1)^{2S+3B+L}$ where $B$ and $L$
are the baryon and lepton numbers and $S$ is spin \cite{Fayet1,Farrar}.
If the lightest supersymetric particles is the tau-sneutrino, it\textquoteright{}s
decay may be realized only via R-parity violation (RPV): $\tilde{\nu}\rightarrow l^{+}l^{'^{-}}$,
$\tilde{\nu}\rightarrow q^{+}q^{'^{-}}$.

If R \textendash{} parity is violating, $e{}^{-}e^{+}\rightarrow\tilde{\nu_{\tau}}\bar{\tilde{\nu}_{\tau}}\rightarrow\mu^{+}\mu^{+}e^{-}e^{-}$
process becomes very important. In this paper $\tilde{\nu_{\tau}}$
pair production at the CLIC with subsequent RPV decays in to $e\mu$
pairs has been investigated. Feynman diagram for tau-sneutrino production
process is shown in Figure 1.

\begin{figure}
\includegraphics[scale=0.7]{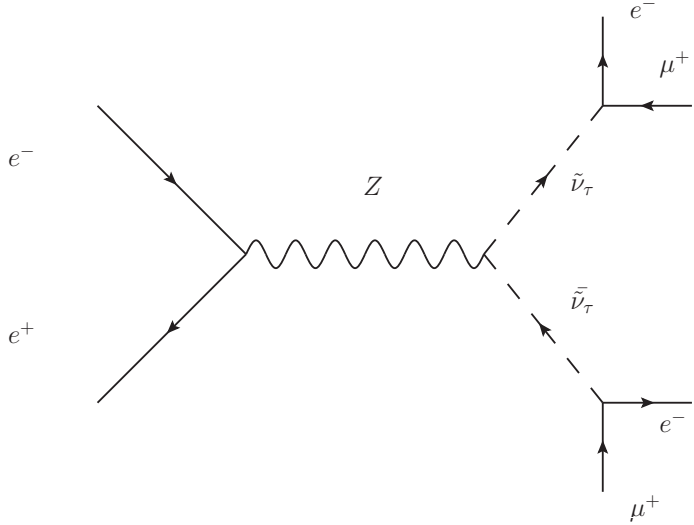}

\caption{Feynman diagram for $e^{+}e^{-}\rightarrow\tilde{\nu_{\tau}}\overline{\tilde{\nu}_{\tau}}\rightarrow\mu^{+}e^{-}\mu^{+}e^{-}$process.}

\end{figure}

The R-parity violation part of the MSSM superpotential is given by
\cite{Barbier}

\begin{equation}
W_{RPV}=\frac{1}{2}\lambda_{ijk}L_{i}L_{j}E_{k}^{c}+\lambda_{ijk}^{'}L_{i}Q_{j}D_{k}^{c}+\frac{1}{2}\lambda_{ijk}^{''}U_{i}^{c}D_{j}^{c}D_{k}^{c}\end{equation}

where $L\,(E)$ is an $SU(2)$ doublet (singlet) lepton superfield
and $Q(U,D)$ is (are) an $SU(2)$ doublet (singlet) quark superfield(s),
and indices $i,\, j,\, k\,=1,\,2,\,3$ denote flavour. The coefficients
$\lambda_{ijk}$ and $\lambda_{ijk}^{''}$ correspond to the lepton
number violating and baryon number violating couplings, respectively. 

RPV interaction Lagrangian responsible for is $\tilde{\nu_{\tau}}\rightarrow\mu^{+}e^{-}$
and $\mu^{-}e^{+}$ decays is given below:

\begin{equation}
L=-\frac{1}{2}\lambda_{312}\widetilde{\nu}_{\tau L}\bar{e}_{R}\mu_{L}-\frac{1}{2}\lambda_{321}\widetilde{\nu}_{\tau L}\bar{\mu}_{R}e_{L}+h.c.\end{equation}

For the numerical calculations we implement the Lagrangian ($2$)
into the CALCHEP MSSM package \cite{CALCHEP}. The cross-section for
pair production of tau-sneutrinos at CLIC with $\sqrt{s}=0.5$ TeV
is shown Figure 2. Initial State Radiation (ISR) and Beamstrahlung
effects at CLIC are calculated with CALCHEP program using beam parameters
given in Table 1 \cite{CLIC,CLIC1}.

\begin{table}
\begin{tabular}{|c|c|c|}
\hline 
Collider Parameters & $\sqrt{s}=0.5$ TeV & $\sqrt{s}=3$ TeV\tabularnewline
\hline
\hline 
$E\,(\sqrt{s})$, TeV & $0.5$ & $3$\tabularnewline
\hline 
$L\,(10^{34}\, cm^{-2}\, s^{-1})$ & $2.3$ & $5.9$\tabularnewline
\hline 
$N\,(10^{10})$ & $0.68$ & $0.372$\tabularnewline
\hline 
$\sigma_{x}\,(nm)$ & $202$ & $45$\tabularnewline
\hline 
$\sigma_{y}\,(nm)$ & $2.3$ & $1$\tabularnewline
\hline 
$\sigma_{z}\,(\mu m)$ & $44$ & $44$\tabularnewline
\hline
\end{tabular}\caption{Main parameters of CLIC. Here N is the number of particles in bunch.
$\sigma_{x}$ and $\sigma_{y}$ are RMS beam sizes at Interaction
Point (IP), $\sigma_{z}$ is the RMS bunch lenght.}

\end{table}

\begin{figure}
\includegraphics[scale=0.7]{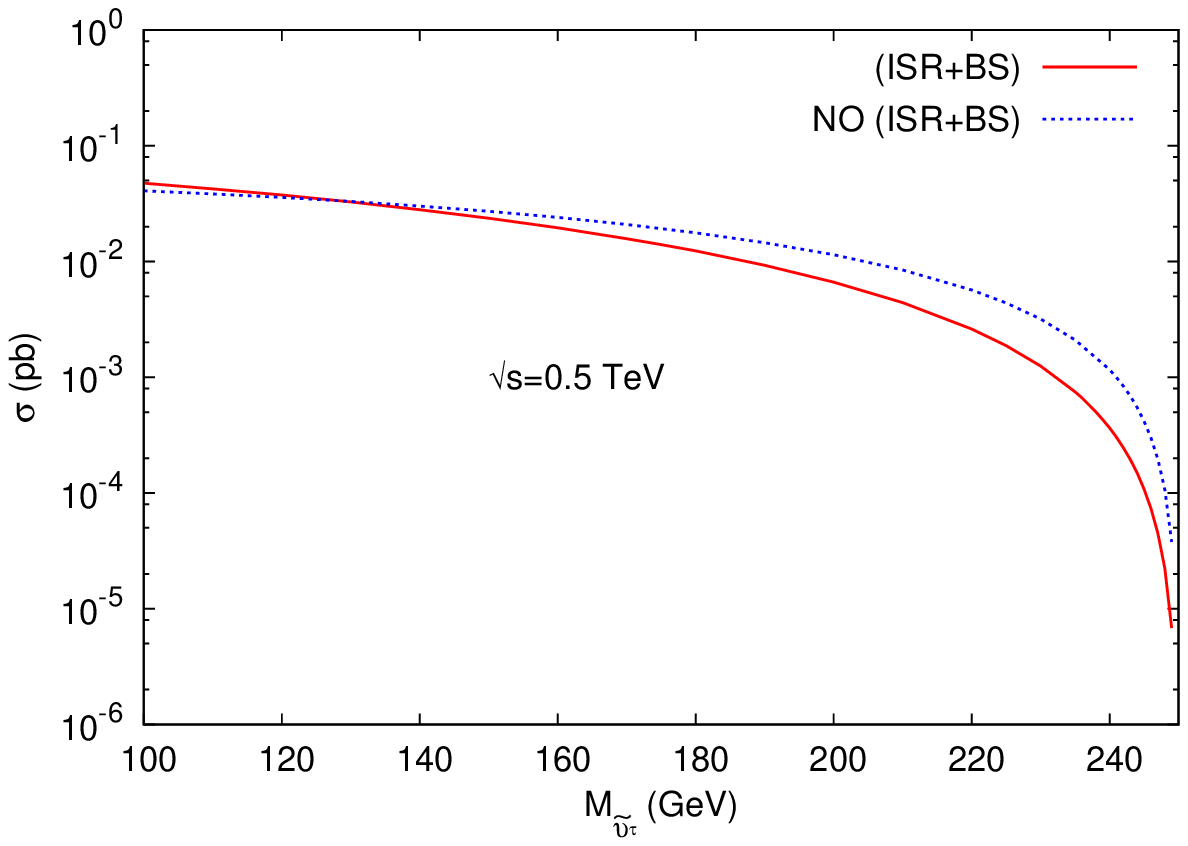}

\caption{Cross sections for $e^{-}e^{+}\rightarrow\widetilde{\nu_{\tau}}$$\overline{\widetilde{\nu_{\tau}}}$
process at CLIC with $\sqrt{s}=0.5$ TeV.}

\end{figure}

\begin{figure}
\includegraphics[scale=0.7]{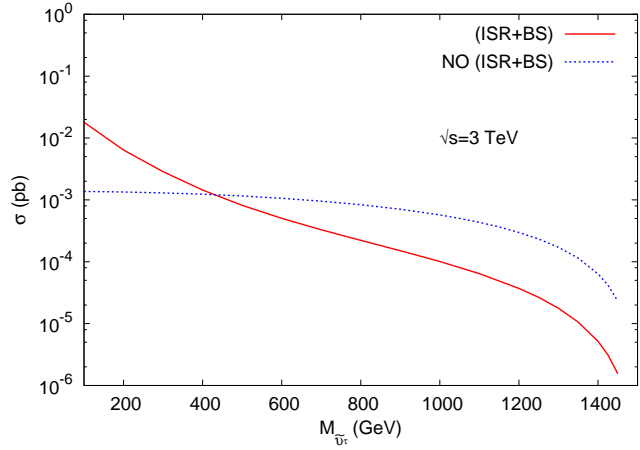}

\caption{Cross sections for $e^{-}e^{+}\rightarrow\widetilde{\nu_{\tau}}\overline{\widetilde{\nu}_{\tau}}$
process at CLIC with $\sqrt{s}=3$ TeV. }

\end{figure}

It is seen from Figure 2, that ISR and BS effects leads to increasing
(decreasing) of the cross-section for $M_{\widetilde{\nu}_{\tau}}$below
(above) 130 GeV. In Figure 3 we present similar calculations for the
CLIC with $\sqrt{s}=3$ TeV. It is seen that ISR and BS effects are
more effective at higher center of mass energy. ISR+BS effects lead
to increasing of the cross sections for $M_{\widetilde{\nu}_{\tau}}<450$
GeV. After $M_{\widetilde{\nu}_{\tau}}=450$ GeV, ISR+BS effects decrease
the cross sections.

We propose $e^{-}e^{+}\rightarrow\widetilde{\nu}_{\tau}\overline{\widetilde{\nu_{\tau}}}\rightarrow(\mu^{+}e^{-})(\mu^{+}e^{-})$
backgroundless process to analysis at CLIC. In order to analyze signal
following basic cuts are applied: $P_{T}>20$ GeV and $|\eta|<2.5$
for the final state electrons and muons. Assuming $\lambda_{321}=\lambda_{312}$
and taking other possible RPV interaction constants to be $0$, which
means $Br\,(\widetilde{\nu}_{\tau}\rightarrow\mu^{+}e^{-})=Br\,(\overline{\widetilde{\nu}_{\tau}}\rightarrow\mu^{+}e^{-})=1/2$,
we obtain cross section values given in Table 2 (3) for CLIC with
$\sqrt{s}=0.5$ TeV ($3$ TeV). Event number given in the last columns
include both $\mu^{+}\mu^{+}e^{-}e^{-}$ and $\mu^{-}\mu^{-}e^{+}e^{+}$
final states.

\begin{table}
\begin{tabular}{|c|c|c|}
\hline 
$M_{\widetilde{\nu}_{\tau}}$ (GeV) & Cross Section (pb) & Event Number ($N$)\tabularnewline
\hline
\hline 
$100$ & $8.00\times10^{-3}$ & $3686$\tabularnewline
\hline 
$120$ & $7.02\times10^{-3}$ & $3230$\tabularnewline
\hline 
$140$ & $5.80\times10^{-3}$ & $2676$\tabularnewline
\hline 
$160$ & $4.25\times10^{-3}$ & $1957$\tabularnewline
\hline 
$180$ & $2.70\times10^{-3}$ & $1269$\tabularnewline
\hline 
$200$ & $1.51\times10^{-3}$ & $696$\tabularnewline
\hline 
$220$ & $6.10\times10^{-4}$ & $279$\tabularnewline
\hline 
$240$ & $9.60\times10^{-5}$ & $44$\tabularnewline
\hline 
$250$ & $1.10\times10^{-5}$ & $5$\tabularnewline
\hline
\end{tabular}

\caption{Cross sections and event number depending on mass of tau-sneutrinos
at $\sqrt{s}=0.5$ TeV with $P_{T}>20$ GeV and $|\eta|<2.5$ cuts.}

\end{table}

\begin{table}
\begin{tabular}{|c|c|c|}
\hline 
$M_{\widetilde{\nu}_{\tau}}$ (GeV) & Cross Section (pb) & Event Number ($N$)\tabularnewline
\hline
\hline 
$100$ & $3.08\times10^{-3}$ & $3629$\tabularnewline
\hline 
$200$ & $1.32\times10^{-3}$ & $1562$\tabularnewline
\hline 
$400$ & $3.35\times10^{-4}$ & $395$\tabularnewline
\hline 
$600$ & $1.21\times10^{-4}$ & $142$\tabularnewline
\hline 
$800$ & $5.35\times10^{-5}$ & $63$\tabularnewline
\hline 
$1000$ & $2.42\times10^{-5}$ & $28$\tabularnewline
\hline 
$1200$ & $8.82\times10^{-6}$ & $10$\tabularnewline
\hline 
$1300$ & $4.21\times10^{-6}$ & $5$\tabularnewline
\hline 
$1400$ & $1.27\times10^{-6}$ & $2$\tabularnewline
\hline
\end{tabular}

\caption{The same as for Table 2 but for $\sqrt{s}=3$ TeV.}

\end{table}

In order to estimate statistical significance we have used following
formula:

\begin{equation}
S=\frac{\sigma_{s}}{\sqrt{\sigma_{s}+\sigma_{B}}}\sqrt{L_{int}}\end{equation}

Here, S is statistical significance, $\sigma_{s}$ is signal cross
sections values, $\sigma_{B}$ is background cross sections and $L_{int}$
is integrated luminosity. We have backgroundless processes therefore
$\sigma_{B}$ is taken zero. From Eq. (3), discovery ($5\sigma$),
observation ($3\sigma$) and exclusion ($2\sigma$) limits for tau-sneutrino
at CLIC with $\sqrt{s}=0.5$ TeV are obtained as follows: achievable
tau-sneutrino mass values are $243$ GeV for discovery, $248$ GeV
for observation and $251$ GeV for exclusion. Corresponding values
for CLIC with $\sqrt{s}=3$ TeV are : $1030$ GeV for discovery, $1225$
GeV for observation and $1325$ for exclusion.

So far the ideal case, namely, maximal possible Branching Ratio (Br)
for the channel under consideration had been analyzed. In more general
case Branching Ratio is less than $1/2$, because of other possible
decay channels. In Figure 4 we present $5\sigma$, $3\sigma$ and
$2\sigma$ plots for $Br\,(\widetilde{\nu}_{\tau}\rightarrow\mu^{+}e^{-})\times Br\,(\overline{\widetilde{\nu_{\tau}}}\rightarrow\mu^{+}e^{-})$
depending on the $\widetilde{\nu}_{\tau}$ mass for CLIC with $\sqrt{s}=0.5$
TeV. One can see from Figure 4 that for $Br\,(\widetilde{\nu}_{\tau}\rightarrow\mu^{+}e^{-})\times Br\,(\overline{\widetilde{\nu}_{\tau}}\rightarrow\mu^{+}e^{-})=2.5\times10^{-3}$
(hundred times smaller than ideal case) $5\sigma$, $3\sigma$ and
$2\sigma$ limits become $150$ GeV, $190$ GeV and $215$ GeV, respectively.
Corresponding plot for $\sqrt{s}=3$ TeV are presented in Figure 5.

\begin{figure}
\includegraphics[scale=0.7]{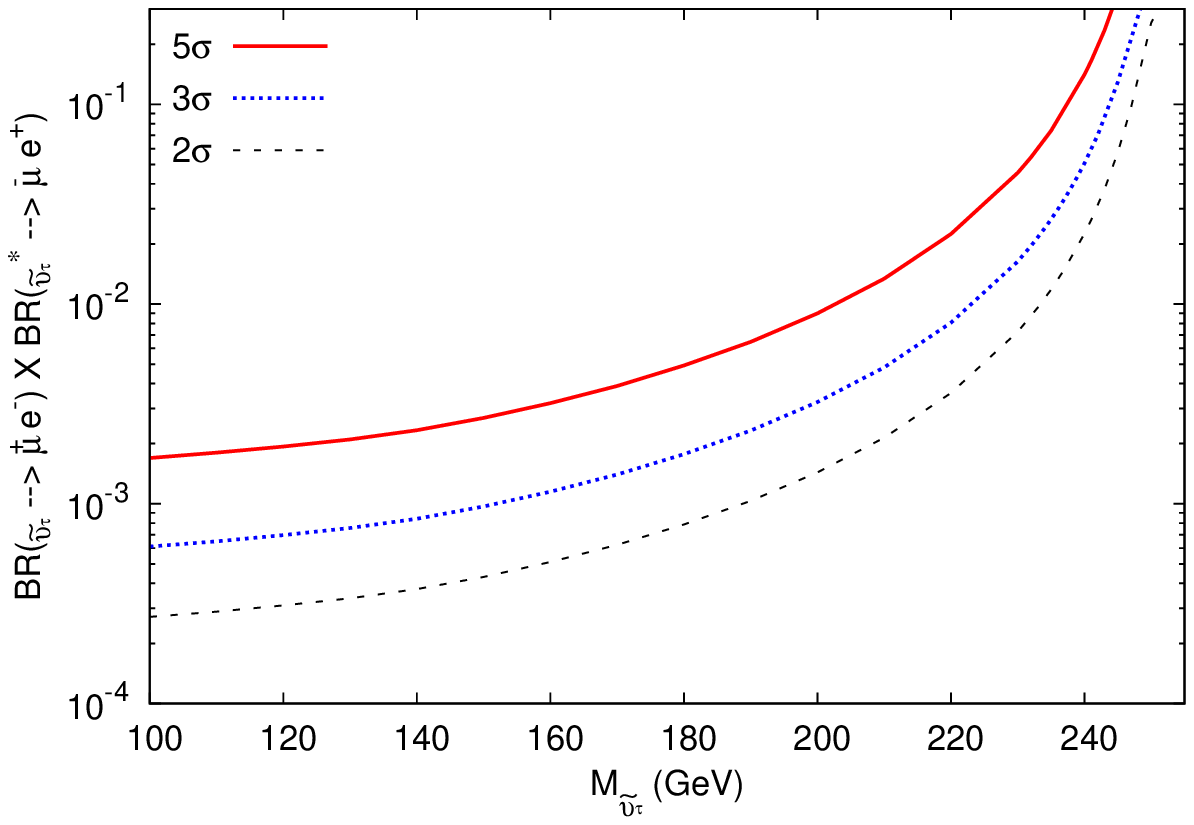}

\caption{Achievable limits for $Br\,(\widetilde{\nu}_{\tau}\rightarrow\mu^{+}e^{-})\times Br\,(\overline{\widetilde{\nu}_{\tau}}\rightarrow\mu^{+}e^{-})$
versus tau-sneutrino mass values at CLIC with $\sqrt{s}=0.5$ TeV. }

\end{figure}

\begin{figure}
\includegraphics[scale=0.7]{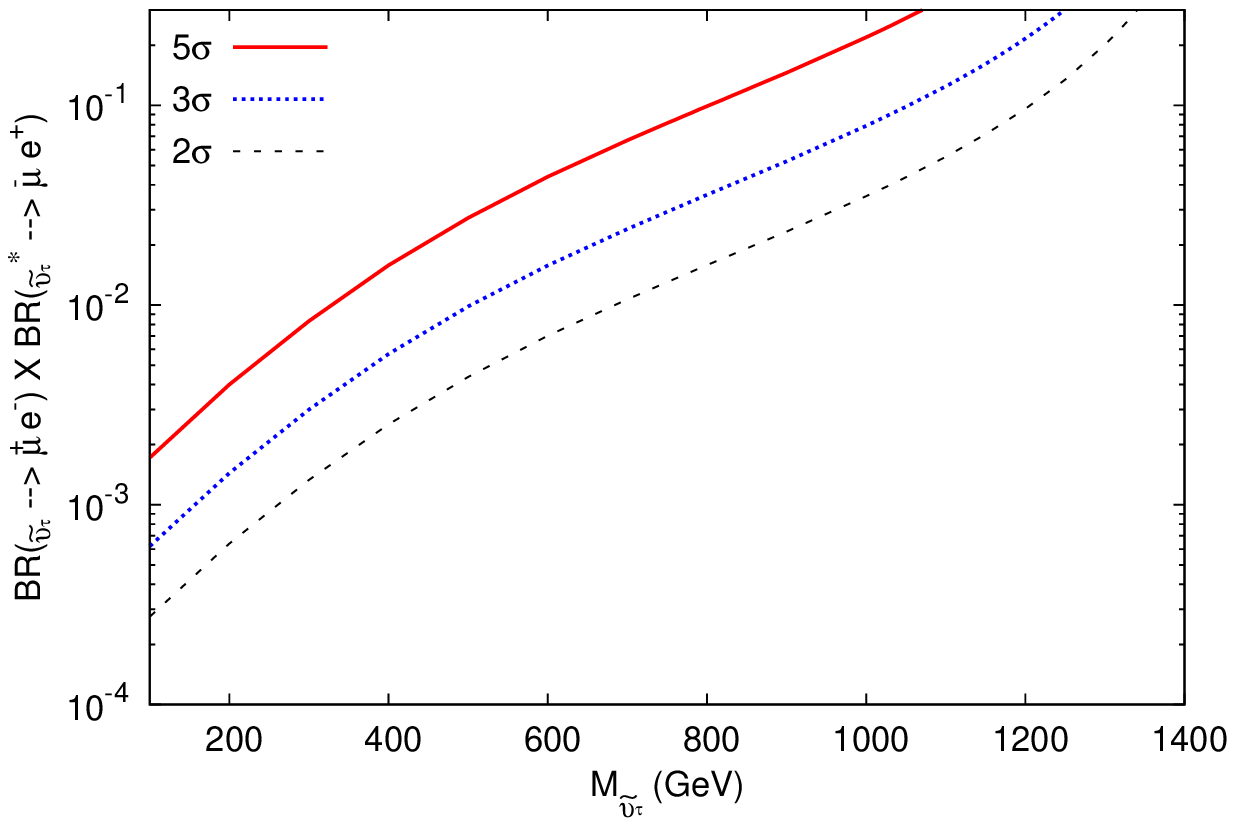}

\caption{Achievable limits for $Br\,(\widetilde{\nu}_{\tau}\rightarrow\mu^{+}e^{-})\times Br\,(\overline{\widetilde{\nu}_{\tau}}\rightarrow\mu^{+}e^{-})$
versus tau-sneutrino mass values at CLIC with $\sqrt{s}=3$ TeV. }

\end{figure}

In the Table 4 (5) discovery, observation and exclusion limits for
$Br\,(\widetilde{\nu}_{\tau}\rightarrow\mu^{+}e^{-})\times Br\,(\overline{\widetilde{\nu}_{\tau}}\rightarrow\mu^{+}e^{-})$
at the CLIC with $\sqrt{s}=0.5$ TeV ($3$ TeV) are given for several
values of the $\widetilde{\nu}_{\tau}$ mass.

\begin{table}
\begin{tabular}{|c|c|c|c|}
\hline 
$M_{\widetilde{\nu}_{\tau}}$, GeV & $5\sigma$ & $3\sigma$ & $2\sigma$\tabularnewline
\hline
\hline 
$120$ & $0.00193$ & $0.00069$ & $0.00031$\tabularnewline
\hline 
$160$ & $0.00319$ & $0.00115$ & $0.00051$\tabularnewline
\hline 
$200$ & $0.00898$ & $0.00323$ & $0.00144$\tabularnewline
\hline 
$240$ & $0.14066$ & $0.05064$ & $0.02251$\tabularnewline
\hline
\end{tabular}

\caption{Achievable limits of $Br\,(\widetilde{\nu}_{\tau}\rightarrow\mu^{+}e^{-})\times Br\,(\overline{\widetilde{\nu}_{\nu}}\rightarrow\mu^{+}e^{-})$
at the CLIC with $\sqrt{s}=0.5$ TeV. }

\end{table}

\begin{table}
\begin{tabular}{|c|c|c|c|}
\hline 
$M_{\widetilde{\nu}_{\tau}}$, GeV & $5\sigma$ & $3\sigma$ & $2\sigma$\tabularnewline
\hline
\hline 
$100$ & $0.00172$ & $0.00062$ & $0.00027$\tabularnewline
\hline 
$400$ & $0.01582$ & $0.00569$ & $0.00253$\tabularnewline
\hline 
$800$ & $0.09899$ & $0.03564$ & $0.01584$\tabularnewline
\hline 
$1100$ & $0.34578$ & $0.12448$ & $0.05532$\tabularnewline
\hline
\end{tabular}

\caption{Achievable limits of $Br\,(\widetilde{\nu}_{\tau}\rightarrow\mu^{+}e^{-})\times Br\,(\overline{\widetilde{\nu}_{\tau}}\rightarrow\mu^{+}e^{-})$
at the CLIC with $\sqrt{s}=3$ TeV. }

\end{table}

In conclusion, the process $e^{+}e^{-}\rightarrow\widetilde{\nu}_{\tau}\overline{\widetilde{\nu}_{\tau}}\rightarrow\mu^{+}\mu^{+}e^{-}e^{-}(\mu^{-}\mu^{-}e^{+}e^{+})$
will provide powerful signature for LSP $\widetilde{\nu}_{\tau}$,
if $Br\,(\widetilde{\nu}_{\tau}\rightarrow\mu^{+}e^{-})$ and $Br\,(\overline{\widetilde{\nu}_{\tau}}\rightarrow\mu^{+}e^{-})$
are sufficiently large.

\end{document}